\RequirePackage{fix-cm}
\documentclass[reqno]{amsproc}
\usepackage{amsmath}
\usepackage{fixltx2e}
\usepackage{microtype}

\usepackage{amsfonts}
\usepackage{amssymb}

\usepackage[citation-order,nobysame]{amsrefs}
\usepackage{xyzbib}

\usepackage{hyperref}
\usepackage{microtype}

\usepackage{pstricks,pst-node,pst-plot}
\numberwithin{equation}{section}
\newcommand\wt{\widetilde}
\newcommand\eps{\varepsilon}

\DeclareMathOperator*\res{\mathrm{Res}}

\newcommand{\rmi}{\mathrm{i}}
\newcommand{\rmd}{\mathrm{d}}

\begin{document}
%%%%%%%%%%%%%%%%%%%%%%%%%%%%%%%%%%%%%%%%%%%%%%%%%%%%%%%%%%%%%%%%%%%%%%%%%%%%%%
% \begin{flushright} 
% File version: \texttt{\currfilename}, \today
% \end{flushright}
% \vspace{1in}
%%%%%%%%%%%%%%%%%%%%%%%%%%%%%%%%%%%%%%%%%%%%%

\title
{Generalized emptiness formation probability 
\\ in the six-vertex model}

\author{F. Colomo}
\address{INFN, Sezione di Firenze, 
Via G. Sansone 1, 50019 Sesto Fiorentino (FI), Italy}
\email{colomo@fi.infn.it}

\author{A. G. Pronko}
\address{Steklov Mathematical 
Institute,
Fontanka 27, 191023 St. Petersburg, Russia}
\email{agp@pdmi.ras.ru}

\author{A. Sportiello}
\address{LIPN, and CNRS, Universit\'e Paris 13, Sorbonne Paris Cit\'e,
99 Av. J.-B. Cl\'ement, 93430 Villetaneuse, France}
\email{Andrea.Sportiello@lipn.univ-paris13.fr}

\begin{abstract}
In the six-vertex model with domain wall boundary conditions, the
emptiness formation probability is the probability that a
rectangular region in the top left corner of the lattice is frozen.
We generalize this notion to the case where the frozen region has
the shape of a generic Young diagram. We derive here a multiple
integral representation for this correlation function.
\end{abstract}

\maketitle
%%%%%%%%%%%%%%%%%%%%%%%%%%%%%%%%%%%%%%%%%%%%%%%%%%%%%%%%%%%%%%%%%%%%%%%%%%%%%%
\section{Introduction}

The six-vertex model with domain wall boundary conditions
\cite{K-82,I-87,ICK-92} attracts interest, in particular, for its
phase separation and limit shape phenomena \cite{KZj-00,Zj-00,Zj-02,
BL-13,RP-06,RS-15,CGP-15,ADSV-15}.  These can be studied
analytically provided that appropriate correlation functions are
known. In this respect, the first results concerned the probabilities
of observing various specific configurations near the boundary, see
\cite{BKZ-02,BPZ-02,FP-04,CP-05c}.

An example of correlation function for configurations away from the
boundaries is the emptiness formation probability (EFP) \cite{CP-07b},
see also \cite{M-11,CP-12}.  This is a nonlocal correlation function,
describing the probability of having the first $s$ consecutive
horizontal edges along a given column, all in a given state. In the
thermodynamic limit the EFP has a simple stepwise behaviour, with the
jump occurring exactly in correspondence of the phase separation curve
(or frozen boundary of the limit shape, or arctic curve) --- a
property that allowed for the determination of the analytic expression
of the arctic curve \cite{CP-08,CP-09,CPZj-10}.

In view of a deeper understanding of limit shape phenomena, and to
address them on wider settings, it is desirable to extend the above
mentioned results to regions of the lattice with more generic shapes.
Some preliminary studies in this direction have already shown the
presence of two important features, namely, the occurrence of a
spatial phase transition in the case of a domain of varying shape
\cite{CP-13,CP-15}, and the fact that, even in the case of generic
domains, arctic curves can be determined from the knowledge of the
corresponding boundary correlation functions \cite{CS-16}.

In the present paper we introduce a nonlocal correlation function that
can provide further advances in these directions.  For the six-vertex
model on a square domain, it describes the probability of having an
$s$-tuple of horizontal edges (one edge per row, for the first $s$
rows, with corresponding column indices forming a weakly ordered
sequence), all in a given state. When the horizontal edges are in the
same column, this function reduces to the EFP.  We thus call it
\emph{generalized emptiness formation probability} (GEFP).

To compute the GEFP, we use the quantum inverse scattering method
\cite{TF-79,KBI-93}. In the derivation, we follow the method developed
in \cite{CP-07b} for calculating the EFP, see also \cite{CP-12}.  Here
we provide the result in the form of a multiple integral
representation, Eq.~\eqref{MIRGEFP}, of which Eq.~(5.17) in
\cite{CP-07b} is a particular case.  The obtained representation is
reminiscent of analogous multiple integral representations for
correlation functions in quantum spin chains
\cite{JM-95,KMST-02,BKS-03,GKS-04,BJMST-06}, asymmetric simple
exclusion process \cite{S-97,BFS-08,TW-08}, and stochastic six-vertex
model \cite{BCG-16}.

%%%%%%%%%%%%%%%%%%%%%%%%%%%%%%%%%%%%%%%%%%%%%%%%%%%%%%%%%%%%%%%%%%%%%%%%%%%
\section{Definition of the GEFP}

In this section we recall the definition of the six-vertex model with
domain wall boundary conditions, and introduce the GEFP.

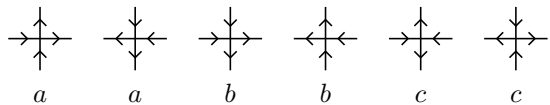
\begin{figure}[b]
\centering
% \documentclass[10pt]{article}
% \usepackage{pstricks,pst-node}
% \pagestyle{empty}
% \begin{document}

%%%%%%%%%%%%%%%%%%%%%%%%%%%%%%%%%%%%%%%%%%%%%%%%%%%%%%%%%%%%%%%%%%%%%%%%%%%%%%
\psset{unit=12pt,linewidth=.05}
\newcommand{\arr}{\lput{:U}{\begin{pspicture}(0,0)
\psline(-.1,.2)(.1,0)(-.1,-.2) \end{pspicture}}}

\begin{pspicture}(0,-1)(17,2)
%\psframe(0,-1)(17,2)
\rput(0,0){
\pcline(0,1)(1,1)\arr \pcline(1,1)(2,1)\arr
\pcline(1,0)(1,1)\arr \pcline(1,1)(1,2)\arr}
\rput(3,0){
\pcline(1,1)(0,1)\arr \pcline(2,1)(1,1)\arr
\pcline(1,2)(1,1)\arr \pcline(1,1)(1,0)\arr}
\rput(6,0){
\pcline(0,1)(1,1)\arr \pcline(1,1)(2,1)\arr
\pcline(1,2)(1,1)\arr \pcline(1,1)(1,0)\arr}
\rput(9,0){
\pcline(2,1)(1,1)\arr \pcline(1,1)(0,1)\arr
\pcline(1,0)(1,1)\arr \pcline(1,1)(1,2)\arr}
\rput(12,0){
\pcline(0,1)(1,1)\arr \pcline(2,1)(1,1)\arr
\pcline(1,1)(1,0)\arr \pcline(1,1)(1,2)\arr}
\rput(15,0){
\pcline(1,1)(2,1)\arr \pcline(1,1)(0,1)\arr
\pcline(1,2)(1,1)\arr \pcline(1,0)(1,1)\arr}
\rput[B](1,-1){$a$}
\rput[B](4,-1){$a$}
\rput[B](7,-1){$b$}
\rput[B](10,-1){$b$}
\rput[B](13,-1){$c$}
\rput[B](16,-1){$c$}
\end{pspicture}

% \end{document}
\caption{The six vertices and their weights.}
\label{fig-sixv}
\end{figure}

We consider the six-vertex model on a square lattice formed by the
intersection of $N$ horizontal and $N$ vertical lines (the $N\times N$
lattice). We use the standard formulation of the model in terms of
configurations of arrows pointing along the edges of the lattice, and
subjected to the ice rule, namely at each lattice site (vertex) there are
exactly two incoming and two outgoing arrows. The six allowed
vertex configurations of arrows and the corresponding Boltzmann
weights are shown in Fig.~\ref{fig-sixv} (see, e.g.,
\cite{LW-72,B-82}, for further details).  The domain wall boundary
conditions  mean that all arrows on the left and right boundaries
are outgoing, while all arrows on the top and bottom boundaries are
incoming.

The partition function is defined as follows:
\begin{equation}
Z_N=\sum_{\mathcal{C}}W(\mathcal{C}),\qquad
W(\mathcal{C})=a^{n_1+n_2}b^{n_3+n_4}c^{n_5+n_6}.
\end{equation}
Here, $\mathcal{C}$ is a configuration of the six-vertex model with
domain wall boundary conditions, and $n_i=n_i(\mathcal{C})$,
$i=1,\dots,6$, is the number of vertices of type $i$ in
$\mathcal{C}$, $\sum_{i}n_i=N^2$.  Let us introduce the parameters
\begin{equation}\label{Dt}
\Delta=\frac{a^2+b^2-c^2}{2ab},\qquad t=\frac{b}{a}.
\end{equation} 
The function $Z_N/(a^{N(N-1)} c^N)$ is a polynomial in $b^2/a^2$ and
$c^2/a^2$, and hence, a polynomial in the parameters $\Delta$ and $t$.
Correlation functions, which can be defined as probabilities of
occurrence of certain arrow configurations, are rational functions in
$\Delta$ and $t$.

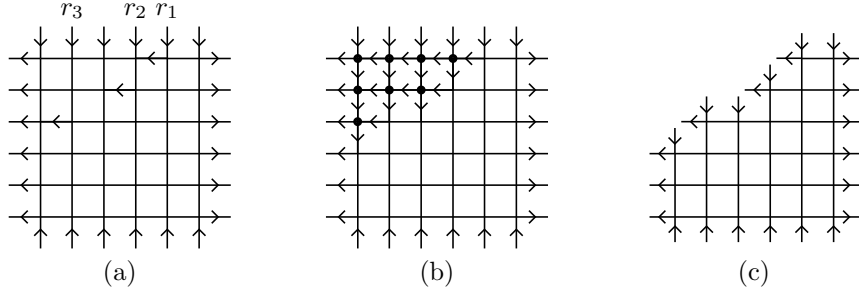
\begin{figure}
\centering
% \documentclass[10pt]{minimal}
% \usepackage{pstricks,pst-node}
% \pagestyle{empty}
% \begin{document}

%%%%%%%%%%%%%%%%%%%%%%%%%%%%%%%%%%%%%%%%%%%%%%%%%%%%%%%%%%%%%%%%%%%%%%%%%%%%%%
\psset{unit=12pt,linewidth=.05}
\newcommand{\arr}{\lput{:U}{\begin{pspicture}(0,0)
\psline(-.1,.2)(.1,0)(-.1,-.2) \end{pspicture}}}

\newcommand{\mydot}{\pscircle[linewidth=.1](0,0){.14}}

\begin{pspicture}(0,-1)(27,7)
%\psframe(0,-1)(27,8)

\multirput(1,0)(1,0){6}{\pcline(0,0)(0,1)\arr}
\multirput(0,1)(0,1){6}{\pcline(6,0)(7,0)\arr}
\multirput(0,1)(0,1){6}{\pcline(1,0)(0,0)\arr}
\multirput(1,0)(1,0){6}{\pcline(0,7)(0,6)\arr}
\multirput(1,1)(1,0){6}{\psline(0,0)(0,5)}
\multirput(1,1)(0,1){6}{\psline(0,0)(5,0)}

\rput(4,6){\pcline(1,0)(0,0)\arr}
\rput(3,5){\pcline(1,0)(0,0)\arr}
\rput(1,4){\pcline(1,0)(0,0)\arr}

\rput(5,7.5){$r_1$}
\rput(4,7.5){$r_2$}
\rput(2,7.5){$r_3$}

\rput[B](3.5,-1){(a)}

\rput(10,0){

\multirput(1,0)(1,0){6}{\pcline(0,0)(0,1)\arr}
\multirput(0,1)(0,1){6}{\pcline(6,0)(7,0)\arr}
\multirput(0,1)(0,1){6}{\pcline(1,0)(0,0)\arr}
\multirput(1,0)(1,0){6}{\pcline(0,7)(0,6)\arr}
\multirput(1,1)(1,0){6}{\psline(0,0)(0,5)}
\multirput(1,1)(0,1){6}{\psline(0,0)(5,0)}

\multirput(4,6)(-1,0){4}{\pcline(1,0)(0,0)\arr}
\multirput(3,5)(-1,0){3}{\pcline(1,0)(0,0)\arr}
\rput(1,4){\pcline(1,0)(0,0)\arr}
\multirput(1,5)(1,0){4}{\pcline(0,1)(0,0)\arr}
\multirput(1,4)(1,0){3}{\pcline(0,1)(0,0)\arr}
\rput(1,3){\pcline(0,1)(0,0)\arr}

\multirput(1,6)(1,0){4}{\mydot}
\multirput(1,5)(1,0){3}{\mydot}
\rput(1,4){\mydot}

%\rput(5,7.5){$r_1$}
%\rput(4,7.5){$r_2$}
%\rput(2,7.5){$r_3$}

\rput[B](3.5,-1){(b)}

}
\rput(20,0){
\multirput(1,0)(1,0){6}{\pcline(0,.2)(0,1)\arr}
\multirput(6,1)(0,1){6}{\pcline(0,0)(.8,0)\arr}
\multirput(0,1)(0,1){3}{\pcline(1,0)(.2,0)\arr}
\multirput(5,6)(1,0){2}{\pcline(0,.8)(0,0)\arr}
\multirput(5,1)(1,0){2}{\psline(0,0)(0,5)}
\multirput(1,1)(0,1){3}{\psline(0,0)(5,0)}
\multirput(1,1)(1,0){4}{\psline(0,0)(0,2)}
\multirput(5,6)(0,-1){3}{\psline(0,0)(1,0)}
\rput(5,5){\psline(0,0)(-1,0)}
\rput(5,4){\psline(0,0)(-3,0)}
\rput(4,3){\psline(0,0)(0,2)}
\multirput(2,3)(1,0){2}{\psline(0,0)(0,1)}
\rput(4,6){\pcline(1,0)(.2,0)\arr}
\rput(3,5){\pcline(1,0)(.2,0)\arr}
\rput(1,4){\pcline(1,0)(.2,0)\arr}
\rput(4,5){\pcline(0,.8)(0,0)\arr}
\multirput(2,4)(1,0){2}{\pcline(0,.8)(0,0)\arr}
\rput(1,3){\pcline(0,.8)(0,0)\arr}

%\rput(5,7.5){$r_1$}
%\rput(4,7.5){$r_2$}
%\rput(2,7.5){$r_3$}

\rput[B](3.5,-1){(c)}
}

\end{pspicture}
%%%%%%%%%%%%%%%%%%%%%%%%%%%%%%%%%%%%%%%%%%%%%%%%%%%%%%%%%%%%%%%%%%%%%%%%%%%%%%

% \end{document}
\caption{(a) The configuration of arrows on the $N\times N$ lattice,
whose probability is described by GEFP; (b) The corresponding frozen
region, made of vertices all of type 2, here marked with dots, has
the shape of the Young diagram $\mu_s$;  (c) The modified domain
obtained by removing the vertices corresponding 
to the Young diagram $\mu_s$ from the top left corner of the
$N\times N$ lattice.  Here, $N=6$, $s=3$, $(r_1,r_3,r_3)=(2,3,5)$,
$\mu_s=(4,3,1)$.}
\label{fig-GEFP}
\end{figure}

We are interested in the probability of observing some specific
configuration of arrows on some given set of edges on the $N \times N$
lattice.  For each edge $e$ of the lattice we define the
characteristic function
\begin{equation}
\chi_e(\mathcal{C})=\left\{\begin{array}{l}
1\quad\mathrm{if\ arrow\ on}\ e\ \mathrm{points\ left\ or\ down}\\
0\quad\mathrm{if\ arrow\ on}\ e\ \mathrm{points\ right\ or\ up.}
\end{array}\right.
\end{equation}
Let us choose $s$ edges, $e_1,\dots, e_s$, $1\leq s\leq N$, with edge
$e_j$, $j=1,\dots,s$, located on the $j$th horizontal line, counting
from the top, and between the $r_j$th and $(r_j+1)$th vertical lines,
counting from the right.  For reasons that will be apparent below, we
require the $r_j$'s to form a weakly increasing sequence,
\begin{equation}\label{r-order}
1\leq r_1\leq r_2 \leq \dots \leq r_s \leq N.
\end{equation}
We denote by $G_{N,s}^{(r_1,\dots,r_s)}$ the probability of observing all
arrows on the horizontal edges $e_1,\dots,e_s$ to be pointing left,
\begin{equation}\label{GEFPdef}
G_{N,s}^{(r_1,\dots,r_s)}=
\frac{1}{Z_N}\sum_{\mathcal{C}}
W(\mathcal{C})\prod_{j=1}^s\chi_{e_j}(\mathcal{C}),
\end{equation}
see Fig.~\ref{fig-GEFP}a.  It is clear that, setting
$r_1=\dots=r_s=r$, the present definition reduces to that of EFP in
\cite{CP-07b}. We thus call this $s$-point correlation function GEFP.

The GEFP satisfies some relations, which follow from the definition
and properties of the model. Due to the ice-rule, if any $r_j\leq j$,
then the probability of the configuration measured by the GEFP vanishes,
and therefore
\begin{equation} \label{property1}
G_{N,s}^{(r_1,\dots,r_s)}>0,
\qquad 
r_j\geq j,\quad j=1,\dots,s.
\end{equation}
On the other hand, if $r_s=N$, then the arrow on the edge $e_s$ may
only point left, due to the domain wall boundary conditions, and 
\begin{equation}\label{property2}
\left. G_{N,s}^{(r_1,\dots,r_s)}\right\vert_{r_s=N}
=G_{N,s-1}^{(r_1,\dots,r_{s-1})},
\end{equation}
that is,  GEFP reduces to that with $s\mapsto s-1$.

We also emphasize that, due to the ice rule and domain wall boundary
conditions, the GEFP equivalently measures the probability that the
vertices at the intersection between the $j$th horizontal line and the
$l_j$th vertical line, $l_j>r_j$, $j=1,\dots,s$, are all of type 2. In
other words, it gives the probability of observing in the top left
corner a frozen region with the shape of a Young diagram
$\mu_s=(m_1,\dots,m_s)$, with rows of length $m_j=N-r_j$,
$j=1,\dots,s$, see Fig.~\ref{fig-GEFP}b.

Finally, we note that the knowledge of the GEFP gives direct access to the
partition function of the six-vertex model on a quite general class of
domains on the square lattice. Specifically, given the set of values
$r_1,\dots,r_j$, $j=1,\dots,s$, let us consider the domain obtained by
removing from the top left corner of the $N\times N$ lattice the
vertices corresponding to the Young diagram $\mu_s$, see
Fig.~\ref{fig-GEFP}c.  In the considered setting, the modified domain
still has boundary conditions of domain wall type, with outgoing
arrows on all horizontal external edges, and incoming arrows on all
vertical external edges, a feature already discussed in \cite{CS-16},
see also \cite{CP-15}. The partition function of the six-vertex model
on the modified domain is exactly given, modulo the factor
$a^{|\mu_s|}/Z_N$, by the GEFP.

%%%%%%%%%%%%%%%%%%%%%%%%%%%%%%%%%%%%%%%%%%%%%%%%%%%%%%%%%%%%%%%%%%%%%%%%%%%
\section{Quantum inverse scattering method calculations}

Here we turn to the calculation of the GEFP. The method developed in
\cite{CP-07b} (see also \cite{CP-12}) in the case of EFP, appears to
be applicable to the GEFP as well. It consists of three steps which we
briefly expose below: first, evaluate the GEFP for the inhomogeneous
model, using the integrability of the six-vertex model, second, take
the homogeneous limit in that expression, and, third, rewrite the
resulting expression as a multiple integral.

The first step is essentially based on commutation relations for
operators entering the quantum monodromy matrix of the six-vertex
model (the Yang-Baxter algebra). These relations make it possible to
derive certain recurrence relations for the GEFP, which, together with
certain initial conditions for the recurrences, can be solved.

The whole procedure of this step is applicable to the inhomogeneous
model, whose weights are parameterized by two sets of spectral
parameters (rapidity variables)
$\boldsymbol{\lambda}:=\{\lambda_1,\dots,\lambda_N\}$ and
$\boldsymbol{\nu}:=\{\nu_1,\dots,\nu_N\}$, and by the crossing parameter
$\eta$ such that the weights of the $(j,k)$-vertex are given by 
\begin{equation}
a_{jk}=a(\lambda_j,\nu_k),\quad 
b_{jk}=b(\lambda_j,\nu_k),\quad 
c_{jk}=c,
\end{equation}
where 
\begin{equation}
a(\lambda,\nu)\equiv\sin(\lambda-\nu+\eta),\quad
b(\lambda,\nu)\equiv\sin(\lambda-\nu-\eta),\quad
c\equiv\sin 2\eta.
\end{equation}
The essential point of this parametrization is that the parameter
$\Delta$ (defined in \eqref{Dt}) is independent of the position of the
vertex, $\Delta=\cos2\eta$. We also denote
\begin{equation}
\varphi(\lambda,\nu)=\frac{c}{a(\lambda,\nu)b(\lambda,\nu)},\quad
d(\lambda,\nu)=\sin(\lambda-\nu),\quad
e(\lambda,\nu)=\sin(\lambda-\nu+2\eta).
\end{equation}
The partition function of the inhomogeneous six-vertex model with
domain wall boundary conditions
is given by the celebrated Izergin-Korepin formula \cite{I-87,ICK-92}:
\begin{equation}\label{IKdet}
Z_N(\boldsymbol{\lambda};\boldsymbol{\nu})
=\frac{\prod_{j,k=1}^{N}a(\lambda_j,\nu_k)b(\lambda_j,\nu_k)}
{\prod_{1\leq j<k \leq N} d(\lambda_k,\lambda_j)d(\nu_j,\nu_k)}
\det[\varphi(\lambda_j,\nu_k)]_{j,k=1,\dots,N},
\end{equation}
Originally, the formula \eqref{IKdet} was proven in \cite{I-87} by
showing that it satisfies certain properties, derived in \cite{K-82},
which completely determine the partition function. Below, we shall
often omit to indicate explicitly the dependence on the sets of
spectral parameters $\boldsymbol{\lambda}$, $\boldsymbol{\nu}$, when
no confusion may arise.

Formula \eqref{IKdet} can also be proven by considering a recurrence
relation valid for generic values of the spectral parameters, which
follows from repeated application of the Yang-Baxter algebra to reduce
the partition function with respect to the weights of a boundary row
(or column).  Relations of this kind were first proposed in
\cite{BPZ-02} to compute one-point boundary correlation functions. In
\cite{CP-07b} it was further observed that these relations can be
recurrently applied $s$ times to compute the EFP.  Here our main
observation is that the same method works also in case of the GEFP.
Denoting
\begin{equation}
\wt G_{N,s}^{(r_1,\dots,r_s)}=
Z_N\,G_{N,s}^{(r_1,\dots,r_s)}
\end{equation}
and applying the very same sequence of steps outlined in
\cite[Section~3]{CP-07b}, we obtain the following recurrence relation
(see also Eq.~(4.1) of that paper):
\begin{multline}\label{Grec}
\wt G_{N,s}^{(r_1,\dots,r_s)}(\boldsymbol{\lambda};\boldsymbol{\nu})
=c\prod_{k=r_1+1}^{N}a(\lambda_k,\nu_1)
\sum_{j=1}^{r_1}
\prod_{\substack{k=1\\ k\ne j}}^{r_1} b(\lambda_k,\nu_1) 
\frac{e(\lambda_k,\lambda_j)}{d(\lambda_k,\lambda_j)} 
\prod_{l=2}^{N} a(\lambda_j,\nu_l)
\\  \times
\wt G_{N-1,s-1}^{(r_2-1,\dots,r_s-1)}
(\boldsymbol{\lambda}\setminus \{\lambda_j\};
\boldsymbol{\nu}\setminus \{\nu_1\}).
\end{multline}
Just as in \cite{CP-07b}, in the derivation of the recurrence relation
\eqref{Grec} it is crucial that the parameters $\lambda_1, \ldots,
\lambda_{r_1}$ are generic, and that
$G_{N,s}^{(r_1,\dots,r_s)}(\boldsymbol{\lambda};\boldsymbol{\nu})$ is
totally symmetric under permutations of these parameters.  The
permutation symmetry  is a consequence of the Yang-Baxter algebra,
provided that  $r_2, \ldots, r_s \geq r_1$.
We now apply the relation $s$ times, thus requiring the conditions
that $\lambda_1, \ldots, \lambda_{r_s}$ are generic, and $r_1 \leq r_2
\leq \ldots \leq r_s$, that is Eq.~\eqref{r-order}. As a result, in
the right-hand side we are left with the partition functions on the
$(N-s)\times (N-s)$ lattice,
\begin{equation}\label{GeqZ}
\wt G_{N-s,0}^{(.)}= Z_{N-s}^{},
\end{equation}
which is known, being given by the expression \eqref{IKdet}.
Thus the relation \eqref{GeqZ} provides the initial condition for the
recurrence relation~\eqref{Grec}.

As a consequence, relation \eqref{Grec} yields an expression for the
GEFP in the form of an $s$-fold sum of $(N-s)\times (N-s)$
determinants. This sum can be regarded as the result of expanding an
$N\times N$ determinant with respect to $s$ columns, that leads to the
following expression:
\begin{multline}\label{inhomGEFP}
G_{N,s}^{(r_1,\dots,r_s)}=
\frac{1}{\det[\varphi(\lambda_j,\nu_k)]_{j,k=1,\dots,N}}
\prod_{j=1}^{s}
\frac{\prod_{k=j+1}^{N}d(\nu_j,\nu_k)
}{\prod_{k=1}^{r_j}a(\lambda_k,\nu_j)\prod_{k=r_j+1}^{N}b(\lambda_k,\nu_j)}
\\ \times
\det \left[
\begin{cases}
\exp\{\lambda_j\partial_{\eps_k}\} & (k\leq s) \\
\varphi(\lambda_j,\nu_k) & (k>s)
\end{cases}
\right]_{j,k=1,\dots,N}
\prod_{1\leq j<k\leq s} 
\frac{a(\eps_j,\nu_k)b(\eps_k,\nu_k)
}{e(\eps_j,\eps_k)}
\\ \times 
\prod_{j=1}^{s}
\frac{\prod_{k=1}^{r_j}e(\lambda_k,\eps_j)
\prod_{k=r_j+1}^{N}d(\lambda_k,\eps_j)
}{\prod_{k=1}^{N}b(\eps_j,\nu_k)}
\Bigg|_{\eps_1,\dots,\eps_s=0}.
\end{multline}
The essential part in this expression is the 
$N\times N$ determinant involving shift operators
$\exp\{\lambda_j\partial_{\eps_k}\}$. Note also the invariance under
the change: $\exp\{\lambda_j\partial_{\eps_k}\}\mapsto
\exp\{(\lambda_j-\lambda)\partial_{\eps_k}\}$ and $\eps_j\mapsto
\eps_j+\lambda$, where $\lambda$ is an arbitrary parameter.

%%%%%%%%%%%%%%%%%%%%%%%%%%%%%%%%%%%%%%%%%%%%%%%%%%%%%%%%%%%%%%%%%%%%%%%%%%%
\section{Homogeneous limit}

The second step consists in evaluating the homogeneous limit,
$\lambda_1,\dots,\lambda_N\to \lambda$ and $\nu_1,\dots,\nu_N\to \nu$,
in the expression \eqref{inhomGEFP} obtained from the quantum inverse
scattering method calculations.

The Boltzmann weights depend on $\lambda-\nu$ only, thus we set
$\nu=0$ without loss of generality.  Due to the above mentioned shift
invariance, the homogeneous limit of the expression \eqref{inhomGEFP}
can be evaluated, see \cite[Section~5.1]{CP-07b},
\begin{multline}\label{homlim}
G_{N,s}^{(r_1,\dots,r_s)}=
\frac{(-1)^{sN}\prod_{j=1}^{s}(N-j)!}
{a^{rs}b^{(N-r)s}
\det\left[\partial_\lambda^{j+k-2}\varphi\right]_{j,k=1,\dots,N}
}
\\ \times
\det \left[
\begin{cases}
\partial_{\eps_k}^{j-1} & (k\leq s) \\
\partial_\lambda^{j+k-s-2}\varphi & (k>s)
\end{cases}
\right]_{j,k=1,\dots,N}
\prod_{1\leq j<k\leq s} 
\frac{\sin(\eps_j+\lambda+\eta)\sin(\eps_k+\lambda-\eta)
}{\sin(\eps_j-\eps_k+2\eta)}
\\ \times 
\prod_{j=1}^{s}
\frac{(\sin\eps_j)^{N-r_j}[\sin(\eps_j-2\eta)]^{r_j}
}{[\sin(\eps_j+\lambda-\eta)]^{N}}
\Bigg|_{\eps_1,\dots,\eps_s=0},
\end{multline}
where 
\begin{equation}
a=a(\lambda)\equiv a(\lambda,0), \quad 
b=b(\lambda)\equiv b(\lambda,0), \quad
\varphi=\varphi(\lambda)\equiv
\varphi(\lambda,0)=c/ab.
\end{equation}

To simplify 
expression \eqref{homlim}, we introduce the polynomials
\begin{equation}
K_n(x)= (-1)^n n! \varphi^{n+1} 
\frac{\det\left[ 
\begin{cases}
x^{j-1} & (k=1) \\
\partial_\lambda^{j+k-3}\varphi & (k\geq 2)
\end{cases}
\right]_{j,k=1,\dots,n+1}
}{\det\left[\partial_\lambda^{j+k-2}\varphi\right]_{j,k=1,\dots,n+1}}.
\end{equation}
In terms of these polynomials the expression \eqref{homlim} can be written 
as an $s\times s$ determinant,
\begin{multline}\label{poly}
G_{N,s}^{(r_1,\dots,r_s)}=
(-1)^{s} 
\det \left[K_{N-s+j-1}(\partial_{\eps_k})\right]_{j,k=1,\dots,s}
\\ \times 
\prod_{1\leq j<k\leq s} 
\frac{1}{\tilde \rho(\eps_j)\rho(\eps_k)[\tilde\omega(\eps_j)\omega(\eps_k)-1]}
\prod_{j=1}^{s}
[\omega(\eps_j)]^{N-r_j} [\rho(\eps_j)]^N
\Bigg|_{\eps_1,\dots,\eps_s=0},
\end{multline}
where 
\begin{equation}\label{omegarho}
\omega(\eps)=\frac{a}{b}\frac{\sin\eps}{\sin(\eps-2\eta)},\qquad
\rho(\eps)=\frac{b}{c}\frac{\sin(\eps-2\eta)}{\sin(\eps+\lambda-\eta)}
=\frac{1}{\omega(\eps)-1},
\end{equation}
and the tilde stands for the transformation $\eta\to -\eta$.  Note
also the relations
\begin{equation}\label{totr}
\tilde \omega(\eps) =\frac{t^2\omega(\eps)}{2t\Delta\omega(\eps)-1},
\qquad 
\tilde \rho(\eps)=\frac{1}{1-\tilde\omega(\eps)},
\end{equation}
where $\Delta$ and $t$ are given by \eqref{Dt}. Relations \eqref{totr}
imply that all functions in \eqref{poly} are expressed rationally in
terms the function $\omega(\eps)$.

%%%%%%%%%%%%%%%%%%%%%%%%%%%%%%%%%%%%%%%%%%%%%%%%%%%%%%%%%%%%%%%%%%%%%%%%%%%
\section{Multiple integral representation}\label{lastsection}

The third and last step consists in rewriting the expression
\eqref{poly} as a multiple contour integral.

As in the case of EFP in \cite{CP-07b}, to express GEFP as an $s$-fold
contour integral, we consider a particular boundary correlation
function for the model on the $N\times N$ lattice, namely $H_N^{(r)}$,
which gives the probability of observing the sole vertex of type 5 in
the first row from the top, exactly at the $r$th site from the right.
In \cite{BPZ-02,CP-05c} it was shown that
\begin{equation}
H_N^{(r)}=
K_{N-1}(\partial_\eps) [\omega(\eps)]^{N-r}[\rho(\eps)]^N\big|_{\eps=0}.
\end{equation}
Below, we will use the corresponding generating function,
\begin{equation}\label{hNz}
h_N(z)=\sum_{r=1}^N H_N^{(r)}z^{r-1}.
\end{equation}
The following identity plays a crucial role in the derivation of an
integral representation for the GEFP: for any function $f(z)$ regular near
the origin,
\begin{equation}\label{Kfint}
K_{N-1}(\partial_\eps) f(\omega(\eps))\big|_{\eps=0}=
\res_{z=0}\frac{(z-1)^{N-1}h_N(z) f(z)}{z^N}.
\end{equation}
The proof is based on the fact that the function $f(z)$, being regular
near the origin, can be treated as a polynomial of degree $N-1$, since
higher powers in $z$ do not contribute to either sides of the identity
(recall that $\omega(\eps)\to 0$ as $\eps\to 0$). The identity
\eqref{Kfint} thus reduces to a linear relation in an $N$ dimensional
vector space.  For details of the proof, see
\cite[Section~5.3]{CP-07b}.

Before applying identity \eqref{Kfint} to  the determinant
representation \eqref{poly}, let us introduce the multivariate
functions
\begin{equation}\label{hNs}
h_{N,s}(z_1,\dots,z_s) =
\frac{
\det\left[z_j^{k-1}(z_j-1)^{s-k}  h_{N-k+1}(z_j)\right]_{j,k=1,\dots, s}
}{
\prod_{1\leq j<k \leq s}^{}(z_j-z_k)
}.
\end{equation}
These functions are symmetric polynomials of degree $N-1$ in each of
their variables, and satisfy the relation
\begin{equation}\label{at1}
\left. h_{N,s}(z_1,\dots,z_s)\right\vert_{z_s=1}
=h_{N,s-1}(z_1,\dots,z_{s-1}).
\end{equation}
These functions are closely related to the partially inhomogeneous
Izergin-Korepin partition function \cite{CP-07b,CP-09}.

Using now the identity \eqref{Kfint} within the determinant
representation \eqref{poly}, and recalling relations \eqref{omegarho}
and \eqref{totr}, we obtain the following multiple integral
representation for the GEFP:
\begin{multline}\label{MIRGEFP}
G_{N,s}^{(r_1,\dots,r_s)}=(-1)^s\oint\cdots\oint
\prod_{j=1}^{s}\frac{[(t^2-2\Delta t)z_j+1]^{s-j}}{z_j^{r_j}(z_j-1)^{s-j+1}}\,
\\ \times
\prod_{1\leq j<k \leq s}^{} \frac{z_j-z_k}{t^2z_jz_k-2\Delta t z_j+1}\,
h_{N,s}(z_1,\dots,z_s)
\,\frac{\rmd^s z}{(2\pi \rmi)^s}.
\end{multline}
Here, the integrations are performed over simple counterclockwise
oriented contours surrounding the origin and no other singularity of
the integrand.

As a simple check of representation \eqref{MIRGEFP}, we note that it
satisfies relations \eqref{property1} and \eqref{property2}.
Concerning the first relation, let us consider the integrand in the
limit $z_1\to 0$, $\dots$, $z_s\to 0$, performed in this order for
convenience, keeping at each stage the contribution of leading order
in the corresponding variable. In this limit the integrand behaves as
$\prod_{j=1}^s z_j^{j-r_j-1}$, and thus the integral vanishes unless
$r_j\geq j$, $j=1,\dots,s$.

Turning to relation \eqref{property2}, we observe that for $r_s\geq
N$, the integrand has no pole at infinity in $z_s$, and thus the
corresponding integration countour can be deformed to enclose the
poles at $z_s=(2\Delta t z_j -1)/(t^2 z_j)$, $j=1,\dots,s-1$, and at
$z_s=1$.  The contribution of each of the first $s-1$ poles vanishes,
due to the property
\begin{equation}\label{assign}
h_{N,s}(z_1,\dots,z_s)\Big|_{z_s= 
\frac{2\Delta t z_j -1}{t^2 z_j}}\propto z_j,
\qquad z_j\to  0,\qquad j=1,\dots,s-1,
\end{equation}
discussed in some detail in appendix A.
The property \eqref{assign}
implies for the integrand of the remaining $(s-1)$-fold integral
the behaviour $z_j^{r_s-r_j}$ as $z_j\to 0$, and thus ensures the
vanishing of the corresponding integration. Therefore, we only need to
evaluate the contribution of the simple pole at $z_s=1$. Using the
relation \eqref{at1}, we reproduce representation \eqref{MIRGEFP}
with $s \mapsto s-1$, and hence get \eqref{property2}.  

The representation \eqref{MIRGEFP} is our main result.  It generalizes
Eq.~(5.17) in \cite{CP-07b} for the EFP.  
Direct comparison shows that the two representations differ
only in the simple replacement of the factor $(z_1\dots z_s)^r$ 
in the denominator of the
integrand, with  $(z_1^{r_1}\dots z_s^{r_s})$,
where the $r_j$'s form a weakly increasing sequence, see
\eqref{r-order}.  Note that, while this
may seem a minor modification of the formula, it actually raises a
problem concerning the symmetrization of the integrand with respect to
permutations of the integration variables $z_1,\dots, z_s$.

It is worth emphasizing that symmetrization of the integrand is
necessary, for example, to perform a saddle-point analysis of the
integral representation \eqref{MIRGEFP} for large $s$, to study the
behaviour of the GEFP in the thermodynamic limit.  
The symmetrization issue can be fixed, for example, for the choice
$r_j=N-s+j$, $j=1,\dots,s$, that gives access to the partition
function of the six-vertex model on a square domain with a cut-off
triangle.  Another interesting example corresponds to the choice
$r_1=l$, $r_2=\dots=r_s=r$, providing the boundary correlation
function for the model in an L-shaped domain, as defined in
\cite{CP-13}, and thus giving access to the corresponding arctic
curve, using the method proposed in \cite{CS-16}.
These special cases will be studied in detail elsewhere.
 
In conclusion, we have introduced the GEFP, a generalization of EFP,
in the six-vertex model with domain wall boundary conditions.  Our
main motivation is that the GEFP it is a powerful tool to study the
six-vertex model on variously shaped portions of the square lattice.
The GEFP can be represented, in particular, as a multiple integral,
that is a particularly suitable form to address its asymptotic
behaviour in the scaling limit. 
We believe this will bring 
further insights on phase separation and limit shape phenomena.

%%%%%%%%%%%%%%%%%%%%%%%%%%%%%%%%%%%%%%%%%%%%%%%%%%%%%%%%%%%%%%%%%%%%%%%%%%%
\section*{Acknowledgments}

This work is partially supported by the EC-FP7 Marie Curie Action
grant IRSES-295234 ``Quantum Integrability, Conformal Field Theory and
Topological Quantum Computation'' (QIFCT).  We thank the Galileo
Galilei Institute for Theoretical Physics (GGI, Florence), research
program on ``Statistical Mechanics, Integrability and Combinatorics''
for hospitality and support at some stage of this work. FC is grateful
to LIPN/Equipe Calin for hospitality at early stage of this work.  AGP
and AS are grateful to INFN, Sezione di Firenze for hospitality and
support at some stage of this work. AGP acknowledges partial support
from the Russian Science Foundation, grant 14-11-00598.

%%%%%%%%%%%%%%%%%%%%%%%%%%%%%%%%%%%%%%%%%%%%%%%%%%%%%%%%%%%%%%%%%%%%%%%%%%%

\appendix
\section{}

Here we prove the property \eqref{assign}. For simplicity, we consider
the case $s=N$ and take $j=1$; for generic $s$ and $j$ the same result
will follow due to the total symmetry with respect to the variables
$z_1,\dots,z_N$, and to the relation \eqref{at1}. We thus need to prove that
the  function
$h_{N,N}(z_1,\dots,z_N)$ has a simple zero at the point
\begin{equation}\label{zszj}
z_N=\frac{2t\Delta z_1-1}{t^2 z_1}, 
\end{equation}
 as $z_1\to 0$. 

In \cite{CP-07b} it was shown the function $h_{N,N}(z_1,\dots,z_N)$
can be expressed in terms of the partition function
$Z_N(\boldsymbol{\lambda})\equiv
Z_N(\boldsymbol{\lambda};\boldsymbol{\nu})|_{\nu_1,\ldots,\nu_N=0}$ as follows:
\begin{equation}\label{hviaZ}
h_{N,N}(z_1,\dots,z_N)=
\frac{Z_N(\boldsymbol{\lambda})}{Z_N}
\prod_{j=1}^{N} \left[\frac{a}{a(\lambda_j)}\right]^{N-1},
\end{equation}
where 
$Z_N\equiv Z_N(\boldsymbol{\lambda})|_{\lambda_1,\ldots,\lambda_N=\lambda}$ 
and
\begin{equation}\label{ujs}
z_j=\frac{a}{b} \frac{b(\lambda_j)}{a(\lambda_j)},\qquad j=1,\ldots,N.
\end{equation}
We recall that $a\equiv a(\lambda)$, $b\equiv b(\lambda)$, 
are related to 
the parameter $t$ used in 
the main text
by $t=b/a$.  Due to the Izergin-Korepin formula,
\begin{equation}\label{Zparhom}
Z_N(\boldsymbol{\lambda})=
\frac{\prod_{j=1}^N [a(\lambda_j)b(\lambda_j)]^N}{
\prod_{n=0}^{N-1} n! 
\prod_{1\leq j<k\leq N}d(\lambda_k,\lambda_j)}
\det\left[\partial_{\lambda_k}^{j-1}\varphi(\lambda_k)\right]_{j,k=1,\dots,N}.
\end{equation}

Let us now consider the relation \eqref{zszj}. In terms of the
rapidities of the inhomogeneous partition function, it implies that
$\lambda_N=\lambda_1-2\eta$, see \eqref{ujs}. Clearly, the function
$Z_N(\boldsymbol{\lambda})|_{\lambda_N=\lambda_1-2\eta}$ is an entire
function in $\lambda_1$ and hence the function
$h_{N,N}(z_1,\dots,z_N)|_{z_N=\frac{2t\Delta z_1-1}{t^2 z_1}}$ is
entire in $z_1$. Furthermore, since
\begin{equation}
\varphi(\pm\eta+\epsilon)
=\frac{1}{\epsilon}\pm\cot2\eta+O(\epsilon),
\qquad \epsilon\to0,
\end{equation}
the first and the last columns of the determinant in \eqref{Zparhom}
coincide as $\lambda_1\to \eta$, and the function
$Z_N(\boldsymbol{\lambda})|_{\lambda_N=\lambda_1-2\eta}$ has a simple zero at
the point $\lambda_1=\eta$. Equivalently,
\begin{equation}
h_{N,N}(z_1,\dots,z_N)\Big|_{z_N=\frac{2t\Delta z_1-1}{t^2 z_1}}\propto z_1,
\qquad z_1\to 0,
\end{equation} 
that is exactly the property \eqref{assign}.

\bibliography{gefp_bib}

\begin{bibdiv}
\begin{biblist}

\bib{K-82}{article}{
      author={Korepin, V.~E.},
       title={{Calculations of norms of Bethe wave functions}},
        date={1982},
     journal={Comm. Math. Phys.},
      volume={86},
       pages={391\ndash 418},
}

\bib{I-87}{article}{
      author={Izergin, A.~G.},
       title={Partition function of the six-vertex model in the finite volume},
        date={1987},
     journal={Sov. Phys. Dokl.},
      volume={32},
       pages={878\ndash 879},
}

\bib{ICK-92}{article}{
      author={Izergin, A.~G.},
      author={Coker, D.~A.},
      author={Korepin, V.~E.},
       title={Determinant formula for the six-vertex model},
        date={1992},
     journal={J. Phys. A},
      volume={25},
       pages={4315\ndash 4334},
}

\bib{KZj-00}{article}{
      author={Korepin, V.~E.},
      author={Zinn-Justin, P.},
       title={Thermodynamic limit of the six-vertex model with domain wall
  boundary conditions},
        date={2000},
     journal={J. Phys. A},
      volume={33},
       pages={7053\ndash 7066},
      eprint={cond-mat/0004250},
}

\bib{Zj-00}{article}{
      author={Zinn-Justin, P.},
       title={Six-vertex model with domain wall boundary conditions and
  one-matrix model},
        date={2000},
     journal={Phys. Rev. E},
      volume={62},
       pages={3411\ndash 3418},
      eprint={math-ph/0005008},
}

\bib{Zj-02}{article}{
      author={Zinn-Justin, P.},
       title={{The influence of boundary conditions in the six-vertex model}},
        date={2002},
      eprint={cond-mat/0205192},
}

\bib{BL-13}{book}{
      author={Bleher, P.},
      author={Liechty, K.},
       title={{Random Matrices and the Six-Vertex Model}},
      series={{CRM monographs series}},
   publisher={American Mathematical Society},
     address={Providence, RI},
        date={2013},
      volume={32},
}

\bib{RP-06}{article}{
      author={Reshetikhin, N.},
      author={Palamarchuk, K.},
       title={{The 6-vertex model with fixed boundary conditions}},
        date={2006},
     journal={PoS},
      volume={Solvay},
       pages={012},
      eprint={1010.5011},
}

\bib{RS-15}{article}{
      author={Reshetikhin, N.},
      author={Sridhar, A.},
       title={{Integrability of limit shapes of the six-vertex model}},
        date={2015},
      eprint={1510.01053},
}

\bib{CGP-15}{article}{
      author={Cugliandolo, L.F.},
      author={Gonnella, G.},
      author={Pelizzola, A.},
       title={Six-vertex model with domain wall boundary conditions in the
  {B}ethe-{P}eierls approximation},
        date={2015},
     journal={J. Stat. Mech. Theory Exp.},
       pages={P06008},
      eprint={1501.00883},
}

\bib{ADSV-15}{article}{
      author={Allegra, N.},
      author={Dubail, J.},
      author={St\'ephan, J.-M.},
      author={Viti, J.},
       title={Inhomogeneous field theory inside the arctic circle},
        date={2015},
      eprint={1512.02872},
}

\bib{BKZ-02}{article}{
      author={Bogoliubov, N.~M.},
      author={Kitaev, A.~V.},
      author={Zvonarev, M.~B.},
       title={Boundary polarization in the six-vertex model},
        date={2002},
     journal={Phys. Rev. E},
      volume={65},
       pages={026126},
      eprint={cond-mat/0107146},
}

\bib{BPZ-02}{article}{
      author={Bogoliubov, N.~M.},
      author={Pronko, A.~G.},
      author={Zvonarev, M.~B.},
       title={Boundary correlation functions of the six-vertex model},
        date={2002},
     journal={J. Phys. A},
      volume={35},
       pages={5525\ndash 5541},
      eprint={math-ph/0203025},
}

\bib{FP-04}{article}{
      author={Foda, O.},
      author={Preston, I.},
       title={On the correlation functions of the domain wall six vertex
  model},
        date={2004},
     journal={J. Stat. Mech.},
      volume={0411},
       pages={P001},
      eprint={math-ph/0409067},
}

\bib{CP-05c}{article}{
      author={Colomo, F.},
      author={Pronko, A.~G.},
       title={On two-point boundary correlations in the six-vertex model with
  domain wall boundary conditions},
        date={2005},
     journal={J. Stat. Mech. Theory Exp.},
       pages={P05010},
      eprint={math-ph/0503049},
}

\bib{CP-07b}{article}{
      author={Colomo, F.},
      author={Pronko, A.~G.},
       title={Emptiness formation probability in the domain-wall six-vertex
  model},
        date={2008},
     journal={Nucl. Phys. B},
      volume={798},
       pages={340\ndash 362},
      eprint={0712.1524},
}

\bib{M-11}{article}{
      author={Motegi, K.},
       title={{Boundary correlation functions of the six and nineteen vertex
  models with domain wall boundary conditions}},
        date={2011},
     journal={Physica A},
      volume={390},
       pages={3337\ndash 3347},
      eprint={1101.0187},
}

\bib{CP-12}{article}{
      author={Colomo, F.},
      author={Pronko, A.~G.},
       title={An approach for calculation of correlation functions in the
  six-vertex model with domain wall boundary conditions},
        date={2012},
     journal={Theor. Math. Phys.},
      volume={171},
       pages={641\ndash 654},
      eprint={1111.4353},
}

\bib{CP-08}{article}{
      author={Colomo, F.},
      author={Pronko, A.~G.},
       title={The limit shape of large alternating-sign matrices},
        date={2010},
     journal={SIAM J. Discrete Math.},
      volume={24},
       pages={1558\ndash 1571},
      eprint={0803.2697},
}

\bib{CP-09}{article}{
      author={Colomo, F.},
      author={Pronko, A.~G.},
       title={The arctic curve of the domain-wall six-vertex model},
        date={2010},
     journal={J. Stat. Phys.},
      volume={138},
       pages={662\ndash 700},
      eprint={0907.1264},
}

\bib{CPZj-10}{article}{
      author={Colomo, F.},
      author={Pronko, A.~G.},
      author={Zinn-Justin, P.},
       title={The arctic curve of the domain-wall six-vertex model in its
  anti-ferroelectric regime},
        date={2010},
     journal={J. Stat. Mech. Theory Exp.},
       pages={L03002},
      eprint={1001.2189},
}

\bib{CP-13}{article}{
      author={Colomo, F.},
      author={Pronko, A.~G.},
       title={Third-order phase transition in random tilings},
        date={2013},
     journal={Phys. Rev. E},
      volume={88},
       pages={042125},
      eprint={1306.6207},
}

\bib{CP-15}{article}{
      author={Colomo, F.},
      author={Pronko, A.~G.},
       title={Thermodynamics of the six-vertex model in an {L}-shaped domain},
        date={2015},
     journal={Comm. Math. Phys.},
      volume={339},
       pages={699\ndash 728},
      eprint={1501.03135},
}

\bib{CS-16}{article}{
      author={Colomo, F.},
      author={Sportiello, A.},
       title={Arctic curve of the six-vertex model on generics domains: the
  {Tangent Method}},
        date={2016},
      eprint={1605.01388},
}

\bib{TF-79}{article}{
      author={Takhtadjan, L.~A.},
      author={Faddeev, L.~D.},
       title={The quantum method of the inverse problem and the {H}eisenberg
  {XYZ} model},
        date={1979},
     journal={Russ. Math. Surveys},
      volume={34},
      number={5},
       pages={11\ndash 68},
}

\bib{KBI-93}{book}{
      author={Korepin, V.~E.},
      author={Bogoliubov, N.~M.},
      author={Izergin, A.~G.},
       title={Quantum inverse scattering method and correlation functions},
   publisher={Cambridge University Press},
     address={Cambridge},
        date={1993},
}

\bib{JM-95}{book}{
      author={Jimbo, M.},
      author={Miwa, T.},
       title={Algebraic analysis of solvable lattice models},
      series={CBMS Regional Conference Series in Mathematics},
   publisher={American Mathematical Society},
     address={Providence, RI},
        date={1995},
      volume={85},
}

\bib{KMST-02}{article}{
      author={Kitanine, N.},
      author={Maillet, J.-M.},
      author={Slavnov, N.~A.},
      author={Terras, V.},
       title={{Spin-spin correlation functions of the XXZ-$1/2$ Heisenberg
  chain in a magnetic field}},
        date={2002},
     journal={Nucl. Phys. B},
      volume={641},
       pages={487\ndash 518},
      eprint={hep-th/0201045},
}

\bib{BKS-03}{article}{
      author={Boos, H.~E.},
      author={Korepin, V.~E.},
      author={Smirnov, F.~A.},
       title={Emptiness formation probability and quantum
  {K}nizhnik-{Z}amolodchikov equation},
        date={2003},
     journal={Nucl. Phys. B},
      volume={658},
       pages={417\ndash 439},
      eprint={hep-th/0209246},
}

\bib{GKS-04}{article}{
      author={G{\"o}hmann, F.},
      author={Kl{\"u}mper, A.},
      author={Seel, A.},
       title={Integral representations for correlation functions of the {XXZ}
  chain at finite temperature},
        date={2004},
     journal={J. Phys A},
      volume={37},
       pages={7625\ndash 7652},
      eprint={hep-th/0405089},
}

\bib{BJMST-06}{article}{
      author={Boos, H.},
      author={Jimbo, M.},
      author={Miwa, T.},
      author={Smirnov, F.},
      author={Takeyama, Y.},
       title={Algebraic representation of correlation functions in integrable
  spin chains},
        date={2006},
     journal={Annales Henri Poincare},
      volume={7},
       pages={1395\ndash 1428},
      eprint={hep-th/0601132},
}

\bib{S-97}{article}{
      author={Sch\"utz, G.~M.},
       title={Exact solution of the master equation for the asymmetric simple
  exclusion process},
        date={1997},
     journal={J. Stat. Phys.},
      volume={88},
       pages={427\ndash 445},
      eprint={cond-mat/9701019},
}

\bib{BFS-08}{article}{
      author={Borodin, A.},
      author={Ferrari, P.~L.},
      author={Sasamoto, T.},
       title={{Large time asymptotics of growth models on space-like paths II:
  PNG and parallel TASEP}},
        date={2008},
     journal={Comm. Math. Phys.},
      volume={283},
       pages={417\ndash 449},
      eprint={0707.4207},
}

\bib{TW-08}{article}{
      author={Tracy, C.~A.},
      author={Widom, H.},
       title={Integral formulas for the asymmetric simple exclusion process},
        date={2008},
     journal={Comm. Math Phys.},
      volume={279},
       pages={815\ndash 844},
      eprint={0704.2633},
}

\bib{BCG-16}{article}{
      author={Borodin, A.},
      author={Corwin, I.},
      author={Gorin, V.},
       title={Stochastic six-vertex model},
        date={2016},
     journal={Duke Math. J.},
      volume={165},
       pages={563\ndash 624},
      eprint={1407.6729},
}

\bib{LW-72}{incollection}{
      author={Lieb, E.~H.},
      author={Wu, F.~Y.},
       title={Two dimensional ferroelectric models},
        date={1972},
   booktitle={Phase transitions and critical phenomena},
      editor={Domb, C.},
      editor={Green, M.~S.},
      volume={1},
   publisher={Academic Press},
     address={London},
       pages={331\ndash 490},
}

\bib{B-82}{book}{
      author={Baxter, R.~J.},
       title={Exactly solved models in statistical mechanics},
   publisher={Academic Press},
     address={San Diego, CA},
        date={1982},
}

\end{biblist}
\end{bibdiv}

\bibliographystyle{plain}

\end{document}